\documentclass[prb,twocolumn,showpacs,preprintnumbers,amsmath,amssymb,superscriptaddress,floatfix,twoside]{revtex4}
\usepackage{times,txfonts}
\usepackage[usenames,dvipsnames]{color}
\usepackage{graphicx}
\usepackage{bm}
\usepackage[bookmarks=true,letterpaper=true,colorlinks=true,urlcolor=blue,linkcolor=blue,citecolor=blue]{hyperref} 
\begin{document}

\title{Tuning MgB$_{\bf{2}}$(0001) surface states}

\author{V. Despoja}
\email{vito@phy.hr}
\affiliation{Donostia International Physics Center (DIPC), P.~Manuel de Lardizabal, E-20018 San Sebasti{\'{a}}n, Spain}
\affiliation{Depto. de F{\'{\i}}sica de Materiales and Centro Mixto CSIC-UPV/EHU, Facultad de Ciencias Quimicas, Universidad del Pa{\'{\i}}s Vasco, Apdo.~1072, E-20018 San Sebasti{\'{a}}n, Spain}
\author{D. J. Mowbray}
\affiliation{Donostia International Physics Center (DIPC), P.~Manuel de Lardizabal, E-20018 San Sebasti{\'{a}}n, Spain}
\affiliation{Nano-Bio Spectroscopy Group and ETSF Scientific Development Centre, Depto.~F{\'{i}}sica de Materiales, Universidad del Pa{\'{\i}}s Vasco, Av.~Tolosa 72, E-20018 San Sebasti{\'{a}}n, Spain}  
\author{V. M. Silkin}
\affiliation{Donostia International Physics Center (DIPC), P.~Manuel de Lardizabal, E-20018 San Sebasti{\'{a}}n, Spain}
\affiliation{Depto. de F{\'{\i}}sica de Materiales and Centro Mixto CSIC-UPV/EHU, Facultad de Ciencias Quimicas, Universidad del Pa{\'{\i}}s Vasco, Apdo.~1072, E-20018 San Sebasti{\'{a}}n, Spain}
\affiliation{IKERBASQUE, Basque Foundation for Science, E-48011 Bilbao, Spain}

\begin{abstract}
Surface state localization and hybridization on B, Mg, and Li terminated MgB$_2$(0001) surfaces are studied within density functional theory (DFT).  
For the B terminated surface we find a low energy B $\sigma_1$ surface state, a $sp_z$ surface state, and B $\sigma_{2}$ and $\sigma_{3}$ quantum well states, which are $90\%$ localized in the topmost B layer. Our results demonstrate that by charging the B atomic layer, either by changing the surface termination or through electro-chemical doping, the B $\sigma$ surface states are shifted down in energy, filled, and delocalize through hybridization as they cross the bulk MgB$_2$ bulk bands.  On the other hand, the $sp_z$ surface state should be shifted up in energy, emptied, and gain an increasingly metallic $s$ character by adding a Mg, Mg$^{+1}$, or Li terminating atomic layer.  These results clearly show both the robust nature of MgB$_2$(0001) surface states, and how their localization and energy range may be tuned by surface termination and charging, with implications for the superconducting and plasmonic behaviour of MgB$_2$.
\end{abstract}


\maketitle
\section{Introduction}

Since 2001 when superconductivity in MgB$_2$ was first discovered \cite{SuperinMgb2}, many calculations of its 
bulk and surface electronic structure have been performed.  In general, these calculations showed a good 
agreement with ARPES experiments, suggesting that MgB$_2$ is not a strongly correlated system, and that 
superconductivity is probably a consequence of some strong electron-phonon coupling mechanism. In other words, MgB$_2$ appears to be an extreme case of an Eliashberg superconductor. Indeed, many studies show strong coupling of $\sigma$ electrons and $e_{2g}$ phonons in the B layer, and a wide superconducting gap in the $\sigma$ 
band has been measured \cite{Exp2}. Also, since the recent fabrication of pristine ultra-thin MgB$_2$ films 
\cite{SurfaceSuper}, surface enhanced superconductivity has become an increasingly active area of research \cite{SurfaceSuper}.

These theoretical investigations greatly contributed to the understanding of the peculiar bulk/surface electronic structure of the seemingly
simple MgB$_2$ crystal. Bulk electronic structure calculations \cite{Bela,Kortus,Mazin,An} show that MgB$_2$ has five main bands, namely 
completely filled $\sigma_1$ B, partially filled $\sigma_2$, $\sigma_{3}$ and $p_z$ B, along with completely unoccupied $s$ Mg bands. 
Bonding within the B layers is mostly covalent ($sp^2$ hybridization) and ionic/metallic between B--Mg--B layers. 

Due to the Mg between the B planes, the nature of in-plane covalent bonding is not typical, as in graphite for example. 
Specifically, the presence of Mg enhances the inter-layer overlap and induces strong hybridization between dangling B $p_z$ 
and Mg $s$ orbitals ($sp_z$ bond). This causes an upward shift of the Mg $s$ and downward shift of the B $p_z$ 
band. In other words, this induces an emptying of the Mg $s$ and filling of the B $p_z$ bands. In this way Mg atoms donate electrons interstitially (between B--Mg layers) and 
become positively ionized. This causes an additional downward shift of the B $\pi$ band relative to the B $\sigma$ bands and 
charge transfer from the $\sigma$ to the $\pi$ band. 

In this way MgB$_2$ has a small hole doping of the higher energy $\sigma_2$ and $\sigma_3$ bands (at the $\Gamma$ point) 
and bonds within the B layers become mixed covalent/metallic. Such hole doping in the B layers enhances the 
superconductivity.  This is similar to the case of intercalated graphites \cite{Intercalated}, where doping 
(inter-plane intercalation) by alkali metals causes the graphite to become superconducting.

The above description is also consistent with charge density distributions calculated in Ref.~\onlinecite{Bela}, where it was seen that Mg atoms are strongly ionized, but electrons are donated interstitially, rather than directly to the
B layers, and inter-layer bonding is much more metallic than ionic. 
Also, it has been shown that less charge is participating in the in-plane B--B bonding in MgB$_2$   compared to the C--C bonding in primitive 
graphite.  This suggests that Mg induces a covalent bonding in the B layer which is not completely saturated and is weakened, i.e.~slightly metallic. 

On the other hand, the electronic structure of the MgB$_2$ surface is much more delicate. 
Due to the weak dispersion of bulk B $\sigma$ bands in the (0001) direction, there exist two wide gaps in the
projected bulk band structure of MgB$_2$ at the $\overline{\Gamma}$ point. This allows for the formation of many different types of surface states.     
Indeed, several studies of the MgB$_2$(0001) surface \cite{Kim,Li,Slava,Servedio} showed the existence of various surface 
and subsurface states, whose positions and existence depend on crystal termination and coverage \cite{Profeta}.    

For the case of a B terminated surface, every bulk band has its own surface state band which follows the upper edge of the corresponding projected bulk band. All these states are mostly localized in the first few B layers. For example, the surface state band for B terminated MgB$_2$ reported in Ref.~\onlinecite{Slava} has $p_z$ symmetry 
and is located on the upper edge of the projected $p_z$ bulk band. It's energy at the $\overline{\Gamma}$ point in the surface Brillouin zone (SPZ) is 
about $-2.7$~eV and its charge density is mostly distributed throughout the first three B layers. 

In the case of the Mg terminated surface, there is one surface state whose energy at the 
$\overline{\Gamma}$ point is about $-2$~eV. This surface state has $sp_z$ symmetry and is mostly localized above the surface plane in the vacuum region (supra-surface).

However, ARPES measurements which followed \cite{Exp1} showed a good agreement with bulk electronic 
structure calculations \cite{Kortus,Mazin,An}, although the peak at $0.5$~eV in the normal photo-emission does not agree with any of the above mentioned surface states. The calculations of Servedio \emph{et al.} \cite{Servedio} 
showed good agreement but are not particularly relevant, as the surface potential barrier was there modeled 
by a step potential, so that the binding energy of the surface state was strongly dependent on the position of the potential step. On the other hand, calculations by Profeta \emph{et al.} \cite{Profeta} showed that the
$sp_z$ surface state at the $\overline{\Gamma}$ point strongly depends on the termination and coverage of the MgB$_2$ 
crystal, indicating that in the experiment \cite{Exp1} the surface termination may have been mixed or had significant surface contamination. Uchiyama \emph{et al.} \cite{Exp2} paid much more attention to surface preparation, and in their measurement the normal photo-emission peak appeared at $-1.0$~eV.   
In Refs.~\onlinecite{Exp3,Exp3b,Exp4} an $18$ monolayer (ML) thick ($18$ ML B and $18$ ML Mg layers) MgB$_2$ film was 
deposited an a Mg(0001) substrate using the Molecular Beam Epitaxy (MBE) co-deposition technique. Following this 
it was determined (using several techniques including XPS, LEED+XRD, X ray absorption) that indeed, there was 
no surface contamination, and ARPES measurements were then performed. The results showed an excellent agreement of the
projected bulk band structure with bulk calculations \cite{Kortus,Mazin}. Agreement with surface band structure calculations 
\cite{Slava} was also quite good. For example, peaks  at about $-1.6$~eV and $-3.2$~eV are measured which correspond to $sp_z$ surface 
states in Mg and B terminated surfaces respectively.  
Measurements in Refs.~\onlinecite{Kortus,Mazin} also suggested that the most stable surface termination is Mg, as was anticipated in Refs.~\onlinecite{Li,Profeta}.

Recent experimental realizations of the MgB$_2$ surface have been brought to an increasingly high level of precision \cite{Exp2,Exp3,Exp3b,Exp4,SurfaceSuper}. 
These experiments were aimed at investigating the fine superconductivity properties of MgB$_2$, e.g., multiple superconductivity gaps \cite{Exp2}, and surface enhanced superconductivity \cite{SurfaceSuper}. However, such clean MgB$_2$ surfaces, with a very well defined electronic 
structure, can also be an excellent starting point for the investigation of both optical and dielectric MgB$_2$ surface properties. It also should be noted that in Ref.~\onlinecite{SurfaceSuper} ultrathin MgB$_2$ films down to $7.5$~nm were fabricated, which is quite close to the thicknesses used in the calculations shown herein.  

Interplay between 2D and 3D electronic structures give rise to the unique dielectric 
\cite{Diel1,Diel2,Eguiluz2002,Eguiluz2006} and extraordinary optical properties \cite{Opt1,Opt2} of the MgB$_2$ crystal. 
On the other hand, because of the plethora of surface, localized and subsurface states, the MgB$_2$ surface dielectric 
and optical properties may be even more interesting. Before providing such theoretical calculations we require an accurate description of the
surface electronic structure.  

For this reason, we will focus here on the investigation of three (B, Mg and Li) terminated 
MgB$_2$ surface electronic structures by comparing the results of three different (two plane wave and real space) {\em ab initio} density functional theory (DFT)
methods.   In so doing, we may demonstrate how by changing the surface termination, charging, and doping the MgB$_2$(0001) we may tune both the localization and energy range of the surface and subsurface states, with potential applications in the areas of plasmonics and superconductivity.
 
The paper is organized as follows. In Sec.~\ref{Compmeth} we describe the {\em ab initio} 
methods used for the surface electronic structure calculations, the description of surface formation energies, and develop criteria to distinguish between surface, subsurface, vacuum, and bulk states. The numerical results are presented in Sec.~\ref{Numres} for B, Mg, and Li terminated surfaces, with results for each termination in separate subsections. These results are then discussed in more detail in Sec.~\ref{Conclus}, where we show how surface termination and electro-chemical doping may be used to tune MgB$_2$(0001) surface states.  This is followed by a concluding section.       

\section{Methodology}

\label{Compmeth}
Structural optimization and electronic structure calculations for each MgB$_2$(0001) surface have been performed by 
combining three different \emph{ab initio} codes.  Specifically, we compare our own pseudopotential based plane wave DFT code (SPPW) and the plane wave self-consistent field DFT code (PWscf) belonging to the Quantum Espresso (QE) package \cite{QE}, with a real-space projector augmented wave function method DFT code (GPAW) \cite{GPAW,GPAWRev}.  

In each case we have employed the Perdew-Zunger local density approximation (LDA) for the exchange correlation (xc)-potential \cite{LDA}. An electronic temperature of $k_B T \approx 0.1$ eV was used to converge the Kohn-Sham wavefunctions, with all energies extrapolated to 0 K.  The electronic density was calculated using a $12\times12\times1$ Monkhorst-Pack special $k$-point mesh, i.e.~by using  $19$ special points in the irreducible Brillouin zone.

In both SPPW and PWscf, LDA based pseudopotentials for Li, Mg and B\cite{pseudopotentials} were used, and the energy spectrum was found to be converged with a $25$~Ry plane wave cutoff. For GPAW we employed a grid spacing $h \approx 0.25$~\AA\ in the MgB$_2$(0001) surface plane, and $h \approx 0.20$~\AA\ normal to the surface, which yielded converged results.

\begin{figure}
\includegraphics[width=0.8\columnwidth]{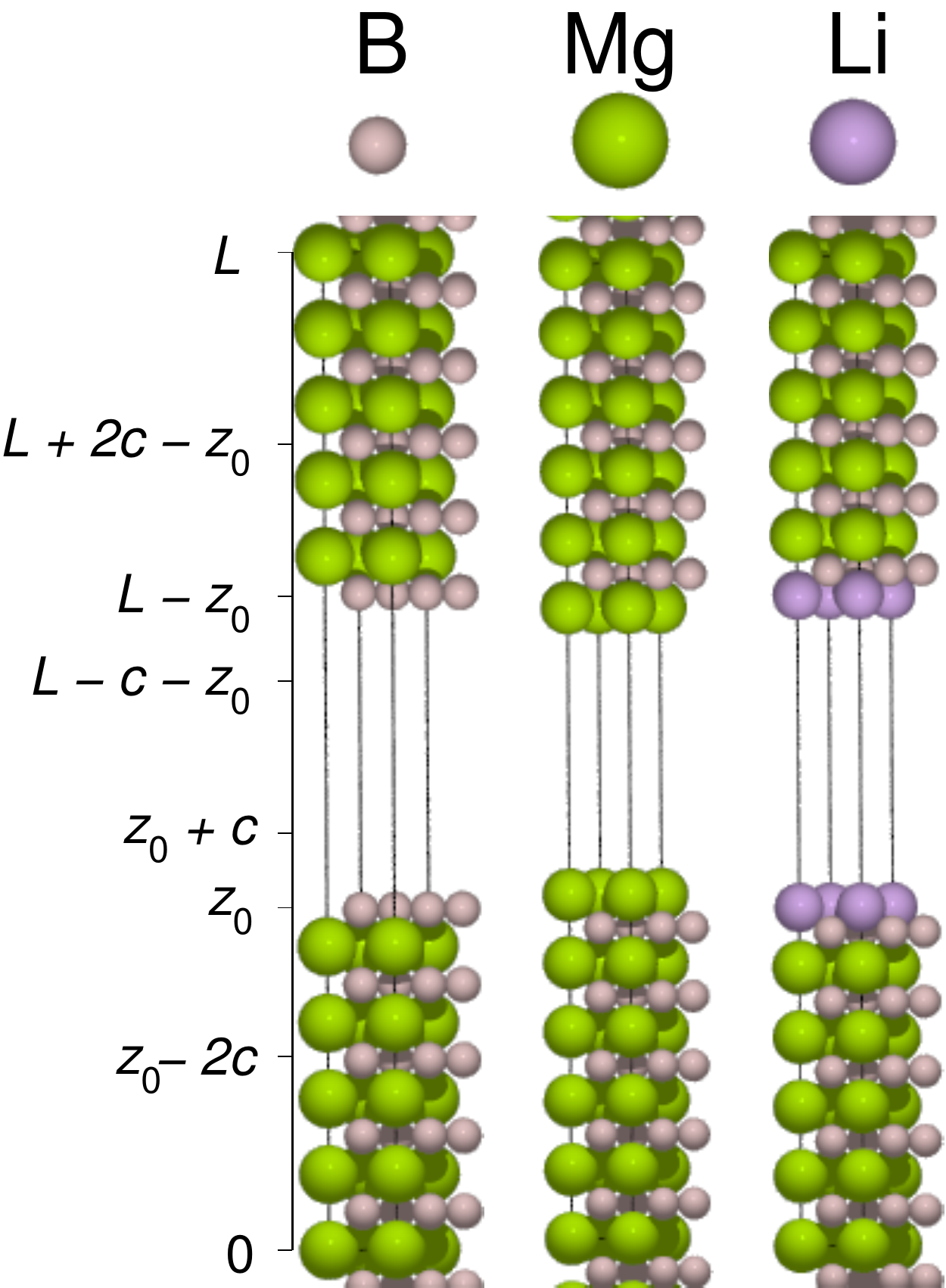}
\caption{Schematic of the structurally optimized MgB$_2$(0001) surface unit cells with B, Mg, and Li termination, repeated twice in each direction.}\label{MgB2_Surfaces}
\end{figure}

As initial atomic coordinates we consider the system to be a hexagonal lattice with lattice parameters $a$ and $c$ taken from the
bulk experimental values, $a\approx5.8317$~a.u.~$\approx 3.086$~\AA~and $c\approx6.6216$~a.u.~$\approx 3.504$~\AA~from Ref.~\onlinecite{SuperinMgb2}, which were previously shown to change by less than 0.6\% upon structural optimization of the MgB$_2$ bulk \cite{Li}. Structural optimization is performed within QE \cite{QE}, and the system is structurally optimized until a maximum force below $0.001$~Ry/a.u.~$\approx 0.026$~eV/\AA~was obtained. 

Schematics of the relaxed MgB$_2$(0001) surface supercell models with B, Mg, and Li termination are shown in Fig.~\ref{MgB2_Surfaces}.  
The B terminated surface was modeled using a supercell consisting of $9$ Mg and $10$ B alternating layers, while the Mg 
terminated surface was obtained by adding Mg layers on top of each surface, with the Li terminated surface formed by replacing the two Mg surface layers with two Li layers. For all terminations four unit cells of vacuum, i.e. $4c \approx 26.4864$~a.u.~$\approx 14.016$~\AA, were used to separate the surfaces. 
This was found to be sufficient to ensure a good description of both surface and bulk states. 

\begin{table}
\caption{GPAW calculated formation energies $\Delta E_f$ for B, Mg, and Li terminated MgB$_2$(0001) surface in eV/\AA$^{2}$ and J/m$^{2}$ relative to bulk MgB$_2$ and the chemical potential of the terminating metal in the bulk for species $X$, $\mu_X$, taken from Ref.~\onlinecite{Kittel}.}\label{GPAW_DEf}
\begin{ruledtabular}
\begin{tabular}{lccc}
MgB$_2$(0001) & B Terminated & Mg Terminated & Li Terminated\\\hline
%$E_{\mathrm{slab}}$ (eV) & -164.31031 &-170.44899& -171.92070\\
$E_{\mathrm{slab}}$ (eV) & -164.31 &-170.45& -171.92\\
%$E_{\mathrm{Bulk}}$ (eV) & -16.95183 &-16.95183 &-16.95183\\
$E_{\mathrm{Bulk}}$ (eV) & -16.95 &-16.95 &-16.95\\
$N_f$ & 9 & 10 & 10\\
$\mu_X$ (eV/atom) & 5.81 & 1.51& 1.63\\
%$\Delta E_f[X]$ (eV/\AA$^2$)& -0.0075 & 0.0351 & -0.03955 \\
$c_\mathrm{B}$, $c_\mathrm{Mg}$, $c_\mathrm{Li}$ & 1, 0, 0 & 0, 1, 0 & 0, -1, 2\\
$\Delta E_f$ (eV/\AA$^2$)& -0.01 & 0.03 & -0.04 \\
$\Delta E_f$ (J/m$^2$)& -0.12 & 0.56 & -0.63 \\
\end{tabular}
\end{ruledtabular}
\end{table}

To estimate the relative stability of the B, Mg, and Li terminated MgB$_2$(0001) surfaces, we consider the surface formation energy $\Delta E_f$, which is defined as the total energy difference from the bulk, and the chemical potential for each of the terminating species, per unit area of the surface \cite{Trends}.  More precisely, 
\begin{eqnarray}
\Delta E_f &\equiv& \frac{E_{\mathrm{Slab}} - N_f E_{\mathrm{Bulk}} - \sum_{X} c_{\mathrm{X}}\mu_{\mathrm{X}}}{2\mathcal{A}},
\end{eqnarray}
where $E_{\mathrm{Slab}}$ is the total energy of the slab, \(\mathcal{A}\approx 29.4524\)~a.u. \(\approx8.2475\)~\AA$^2$ is the surface area of the unit cell, $N_f$ is the number of MgB$_2$ bulk layers per unit cell, \(c_X\) is the number of atoms added or removed from the slab to obtain a given surface termination and \(\mu_X\) is the chemical potential for species $X \in \{\mathrm{B,Mg,Li}\}$, as taken from Ref.~\onlinecite{Kittel}. Values for all parameters are provided in Table~\ref{GPAW_DEf}.

To differentiate between surface states, vacuum states, subsurface states, and bulk states, we first consider the wave function's projected density in the $z$-direction normal to the surface \(\varrho_{n,\mathbf{k}}(z)\). This is defined as
\begin{eqnarray}
\varrho_{n,\mathbf{k}}(z) &\equiv& \iint_{\mathcal{S}} \left|\psi_{n,\mathbf{k}}(x,y,z)\right|^2 dx dy,
\end{eqnarray}
where $\{x,y\}$ are surface coordinates, $\mathcal{S}$ is the surface area of the unit cell, and \(\psi_{n,\mathbf{k}}\) is the \(n^{\mathrm{th}}\) Kohn- Sham wave function at $k$-point $\mathbf{k}$.  We then apply the criterion that the wave function's weight on the surface, \(s_{n,\mathbf{k}}\), surpasses a certain threshold, where we define the surface region as two bulk unit cells below the surface layer, and one unit cell of vacuum above the surface layer.  In more detail, we define
\begin{eqnarray}
s_{n,\mathbf{k}} &=& \int_{z_0-2c}^{z_0+c}\varrho_{n,\mathbf{k}}(z)dz + \int_{L-z_0-c}^{L-z_0+2c}\varrho_{n,\mathbf{k}}(z)dz,
\end{eqnarray}
where $z_0$ is the surface layer coordinate and $L$ is the length of the unit cell in the $z$-direction, as depicted in Fig.~\ref{MgB2_Surfaces}.  States which have more than two thirds of their weight in the surface region, i.e.~$s_{n,\mathbf{k}} \gtrsim 0.66$, are then considered to be surface states.

In order to further distinguish between subsurface states which penetrate deep into the crystal, surface resonant states, and bulk states, we increased the thickness of the supercell, by adding 
six bulk unit cells in the center of the slab.  The resulting structure was then relaxed within QE employing a coarser $k$-point mesh of $8\times8\times1$, to obtain the ground-state geometry.  

To understand how surface states disperse through the SBZ and hybridize with bulk states, we consider the overlap integrals between the Kohn-Sham wave functions at neighboring $k$-points, separated by $\Delta{\mathbf{k}}$.  In real space this is defined simply as
\begin{eqnarray}
\langle\psi_{n,\mathbf{k}}|\psi_{n',\mathbf{k}+\Delta\mathbf{k}}\rangle_{\mathcal{V}}&=&
\int_{\mathcal{V}} \psi_{n,\mathbf{k}}^*(\mathbf{r})\psi_{n',\mathbf{k}+\Delta\mathbf{k}}(\mathbf{r})d\mathbf{r},
\end{eqnarray}
where $\mathbf{r} = \{x,y,z\}$ is the real-space coordinate, and $\mathcal{V}$ is the volume of the unit cell.  Then \(\psi_{n,\mathbf{k}}\rightarrow \psi_{n',\mathbf{k}+\Delta\mathbf{k}}\) for \(n'\) which maximizes \(\left|\langle\psi_{n,\mathbf{k}}|\psi_{n',\mathbf{k}+\Delta\mathbf{k}}\rangle_{\mathcal{V}}\right|\).  In this way we may distinguish between band crossings and avoided crossings in the band structure, and trace the surface states throughout the SBZ.  

\section{Results}
\label{Numres}

\begin{figure*}
\includegraphics[width=0.7\textwidth]{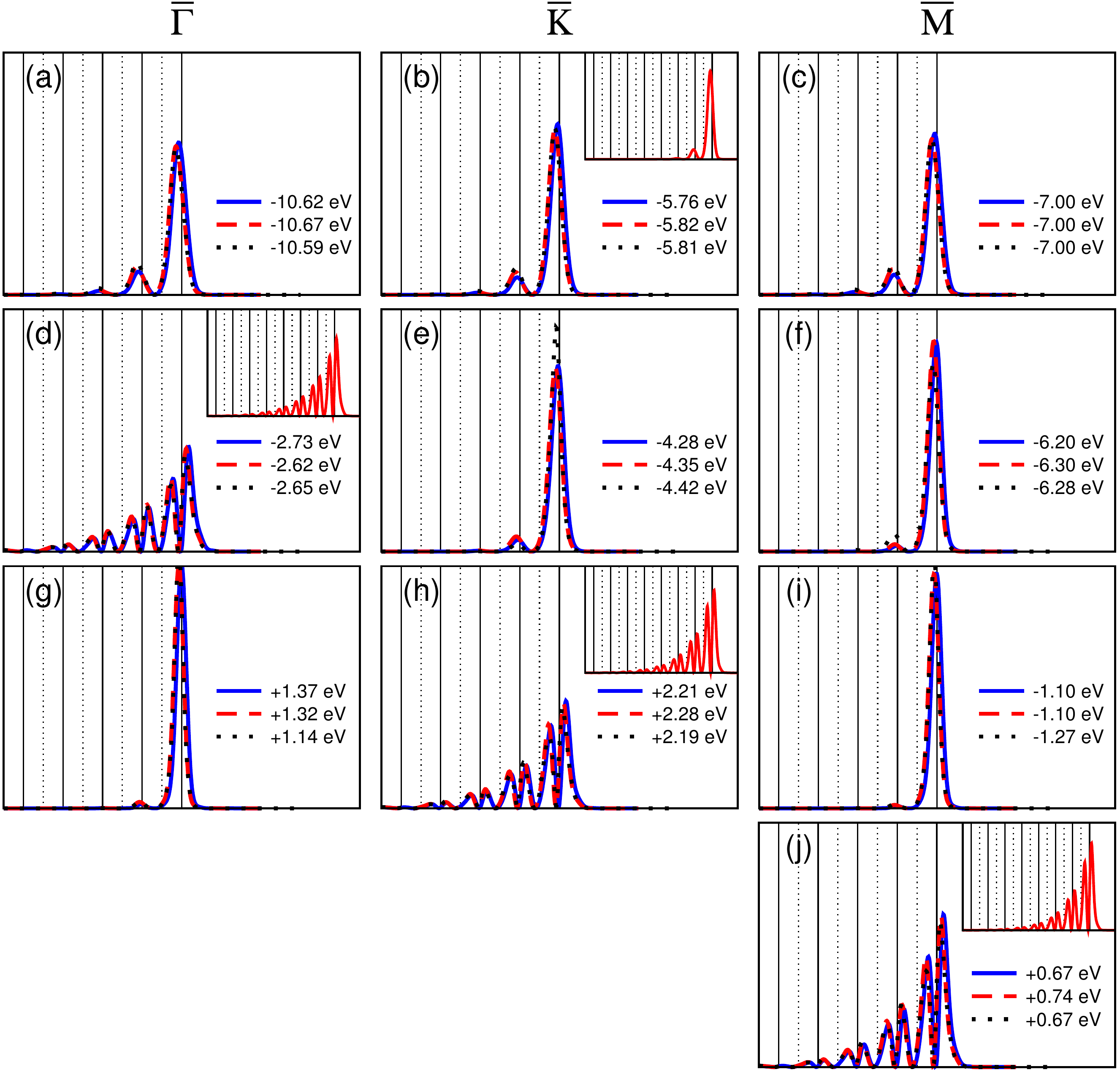}
\caption{(Color online) Projected density of surface and subsurface states \(\varrho_{n,\mathbf{k}}(z)\) in the B terminated MgB$_2$(0001) 
surface versus position normal to the surface plane $z$, and Kohn-Sham eigenenergies $\varepsilon_{n,\mathbf{k}}$ relative to the Fermi level $\varepsilon_F$.  Results from plane wave PWscf calculations before ({\color{blue}{\bf{---}}}) and after ({\color{red}{\bf{-- --}}}) structural optimization are compared with real-space GPAW calculations for the relaxed structure ($\cdots$). Solid and dotted vertical lines denote positions of the B and Mg atomic layers, respectively. Inserts show $\varrho_{n,\mathbf{k}}(z)$ from plane wave PWscf calculations for a relaxed extended supercell model with six bulk unit cells added in the center of the slab. }
\label{Fig2}
\end{figure*} 

\begin{figure}%[h]
%\vspace{2cm}
\includegraphics[width=\columnwidth]{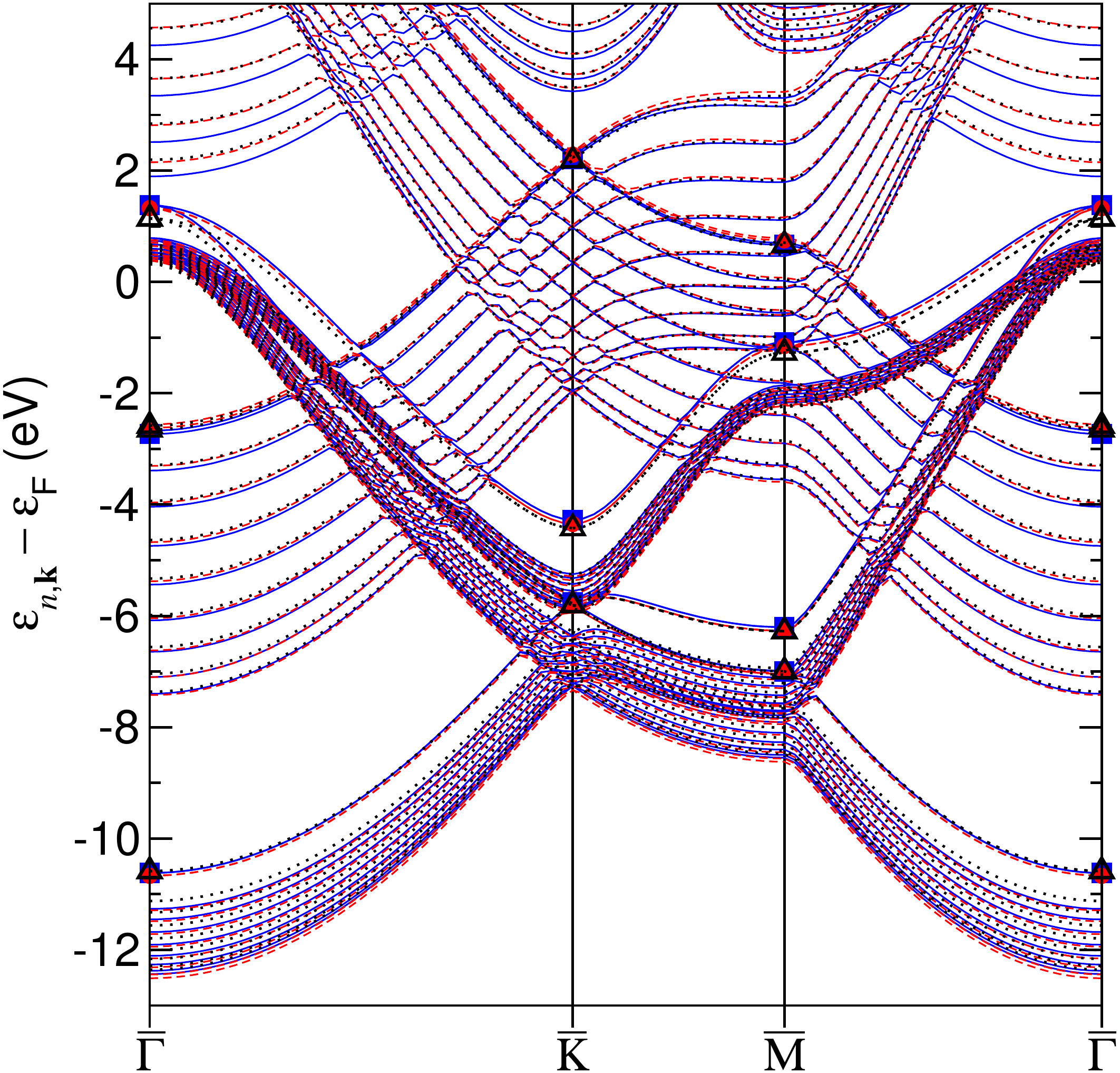}
\caption{(Color online) Band structure (lines) and energies of the surface and subsurface states shown in Fig.~\ref{Fig2} (symbols) for the B terminated 
MgB$_2$(0001) surface in eV relative to the Fermi level $\varepsilon_F$. Results from plane wave PWscf calculations before ({\color{blue}{\bf{---}}}; {\color{blue}{$\blacksquare$}}) and after ({\color{red}{\bf{-- --}}}; {\color{red}{$\medbullet$}}) structural optimization are compared with real-space GPAW calculations for the relaxed structure ($\cdots$; $\vartriangle$).}
\label{Fig1}
\end{figure}

\begin{figure}%[h]
%\vspace{2cm}
\includegraphics[width=\columnwidth]{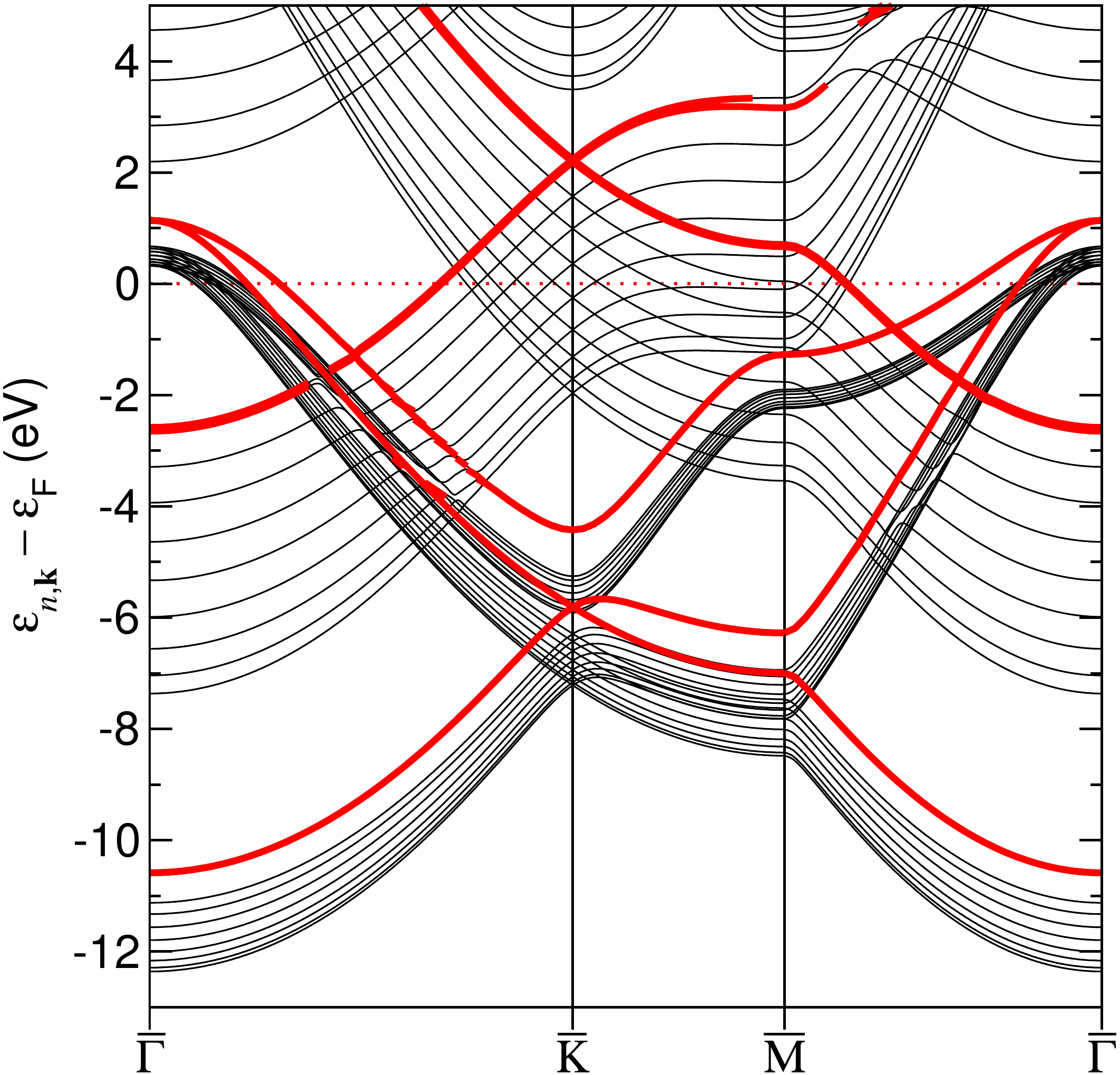}
\caption{(Color online) GPAW overlap-based band structure of relaxed B terminated 
MgB$_2$(0001) surface, with \(\psi_{n,\mathbf{k}}\rightarrow\psi_{n',\mathbf{k}+\Delta\mathbf{k}}\) for \(n'\) which maximizes \(\left|\langle\psi_{n,\mathbf{k}}|\psi_{n',\mathbf{k}+\Delta\mathbf{k}}\rangle_{\mathcal{V}}\right|\).  Surface states with more than two thirds of their weight in the first vacuum and first two bulk layers,  $s_{n,\mathbf{k}} \gtrsim 0.66$, are shown separately ({\color{red}{\bf{---}}}).}
\label{Fig_GPAW_B}
\end{figure}

The DFT estimates of the surface formation energy $\Delta E_f$ shown in Table~\ref{GPAW_DEf} are quite small ($|\Delta E_f| \lesssim 0.05$~eV/\AA$^2$), which suggests that each of the MgB$_2$(0001) surface terminations considered should be stable experimentally. This is clear when one compares with the much higher surface formation energies obtained for other materials, such as metal oxides, which are known to be stable experimentally \cite{Trends}.  In particular, we find it should be thermodynamically preferred to replace a Mg surface termination by a Li termination, so that surface doping by Li is quite feasible.  It should be noted that our choice of bulk metallic B, Mg, and Li for the chemical potential's shown in Table~\ref{GPAW_DEf} has a significant influence on the resulting formation energies obtained.  However, the main conclusion that each surface termination should be obtainable under particular experimental conditions should still hold.

The projection of the bulk band structure in a direction perpendicular to the crystal surface ($\Gamma\rightarrow\mathrm{A}$) is a fingerprint of the particular 
crystal structure. The same is true for a MgB$_2$(0001) surface, for which the projected band structure is actually quite simple. Due to the weak B--B layer overlap, there are three rather narrow B $\sigma$ bands ($\sigma_1$, $\sigma_2$, and $\sigma_3$), while the overlap between $p_z$ orbitals in the B layers and $s$ orbitals in Mg layers yields a wider $\pi$ band. 
The $\pi$ band is then separated from $\sigma$ bands by two wide gaps, which are noticeable in Figs.~\ref{Fig1}, \ref{Fig3} 
and \ref{Fig5}. 

Since the surface band structure of B and Mg terminated MgB$_2$(0001) surfaces was previously studied in Refs.~\onlinecite{Slava,Profeta}, we will focus here on the detailed classification of the surface and subsurface states and surface state 
resonances in Mg, B and Li terminated MgB$_2$(0001).  This is accomplished by combining both pseudopotential-based plane wave methods (SPPW, PWScf) and a real space projector augmented wave function method (GPAW) to test the robustness of the surface states, and geometry optimization to show how 
surface relaxation can modify them.       

\subsection{B-terminated MgB$_{\bf 2}$(0001) surface}

\begin{table}
\caption{Change in B terminated Mg--B inter-layer separation $\Delta_{i,j}$ between the $i^{\mathrm{th}}$ and $j^{\mathrm{th}}$ layers from the surface,  relative to the bulk MgB$_2$ experimental geometry.}\label{Table1}
\begin{ruledtabular}
\begin{tabular}{lrrrr}
&\multicolumn{2}{c}{PWscf LDA\footnote{This Work}}&\multicolumn{2}{c}{VASP PBE\footnote{Ref.~\onlinecite{Li}}}
\\\hline
$\Delta_{0,1}$[B--Mg]& $-4.4\%$&-7.8 pm& $-2.1\%$& -3.7 pm\\
$\Delta_{1,2}$[Mg--B]& $ 0.4\%$& 0.75 pm& $ 2.0\%$&  3.5 pm\\
$\Delta_{2,3}$[B--Mg]& $-1.1\%$&-2.0 pm& $ 0.9\%$&  1.6 pm\\
$\Delta_{3,4}$[Mg--B]& $-1.7\%$&-3.1 pm& $-1.8\%$& -3.2 pm\\
$\Delta_{4,5}$[B--Mg]& $-1.3\%$&-2.6 pm&  $0\%$& 0.0 pm\\
$\Delta_{5,6}$[Mg--B]& $-1.3\%$&-2.3 pm&---&---\\
$\Delta_{6,7}$[B--Mg]& $-1.5\%$&-2.7 pm&---&---\\
$\Delta_{7,8}$[Mg--B]& $-1.3\%$&-2.3 pm&---&---\\
$\Delta_{8,9}$[B--Mg]& $-0.75\%$&-1.4 pm&---&---\\
\end{tabular}
\end{ruledtabular}
\end{table}

To understand how movement of the atomic planes in the surface region influences the energy and character of 
the surface and subsurface states, the B terminated surface  electronic structure has been calculated both with and without 
structural optimization. The relative changes in the (B--Mg) inter-layer separation compared to the bulk 
values, $\Delta_{ij}$, are provided in Table~\ref{Table1}.

We find that the relative change in the inter-layer separation is quite small, in agreement with previous plane wave calculations \cite{Li} using the PBE xc-functional \cite{PBE} where only the top four atomic layers were relaxed. 
This rigidity of the structure is reflected in the band structure 
calculations shown in Fig.~\ref{Fig1}.  We find that the band structures with and without structural optimization do not differ significantly for the B terminated surface.  

The energies of the surface and subsurface states at the $\overline{\Gamma}$, $\overline{\text{K}}$and $\overline{\text{M}}$ points for the structurally unoptimized and optimized surfaces are also shown in Fig.~\ref{Fig1}.    
We find that the B terminated surface has several surface and subsurface states, consisting of three localized bands which are well separated from their corresponding bulk 
bands. For these localized bands we plot $\varrho_{n,\mathbf{k}}$ in Fig.~\ref{Fig2} at $\overline{\Gamma}$,  $\overline{\text{K}}$ and $\overline{\text{M}}$ points. 
The deepest subsurface state's band (Fig.~\ref{Fig2}(a,b,c)) consists of B $\sigma_1$ orbitals, and is localized mainly in 
the first and slightly in the second B layer. Specifically, we find for this band $s_{n,\mathbf{k}} \gtrsim 0.66$ throughout the SBZ, as shown in Fig.~\ref{Fig_GPAW_B}.  In fact, the B $\sigma_1$ surface band may be described semi-quantitatively by a $+0.6$~eV shift of the top of the B $\sigma_1$ bulk bands.

To differentiate between localized states, surface resonances, and bulk states, we recalculated the electronic structure for a thicker slab with six additional MgB$_2$ bulk layers. These results are shown as insets in the upper right corners of the $\varrho_{n,\mathbf{k}}(z)$ plots in Fig.~\ref{Fig2}. For the larger slab we clearly see in Fig.~\ref{Fig2}(b) that the B $\sigma_1$ state is indeed a surface state localized on the topmost B atomic layer. 

The second group of surface states (Fig.~\ref{Fig2}(d,h,j)) are a combination 
of Mg $s$ and B $p_z$ orbitals. These $sp_z$ surface states are mainly localized above the surface and between B and Mg layers. Figure~\ref{Fig_GPAW_B} shows that the $sp_z$ surface state is clearly recognizable as it disperses through the SBZ, although some hybridization occurs when crossing the narrow B $\sigma_{2,3}$ band.  As with the B $\sigma_1$ band, the $sp_z$ surface band is well described by a $+0.6$~eV shift of the top of the respective bulk band, in this case the $\pi$ bands.  We also find for B termination the $sp_z$ states decay slowly into bulk, which is also seen for the extended slab shown as insets in Fig.~\ref{Fig2}(d,h,j).  

The third group of surface states (Fig.~\ref{Fig2} (g,e,i)) are B 
$\sigma$ states which are more than $90\%$ localized in the topmost B layer. From Fig.~\ref{Fig_GPAW_B} we see that the B $\sigma_{2,3}$ surface states may be clearly traced throughout the SBZ, at about 0.6~eV above their respective B $\sigma_{2,3}$ bulk bands.  Thus, Fig.~\ref{Fig_GPAW_B} shows that for the B terminated MgB$_2$(0001) surface the B $\sigma_1$, $sp_z$, and B $\sigma_{2,3}$ surface bands are near quantitatively described by rigidly shifting the top of the respective bulk bands up in energy by 0.6~eV.

\subsection{Mg-terminated MgB$_{\bf 2}$(0001) surface}

\begin{figure*}
\includegraphics[width=0.7\textwidth]{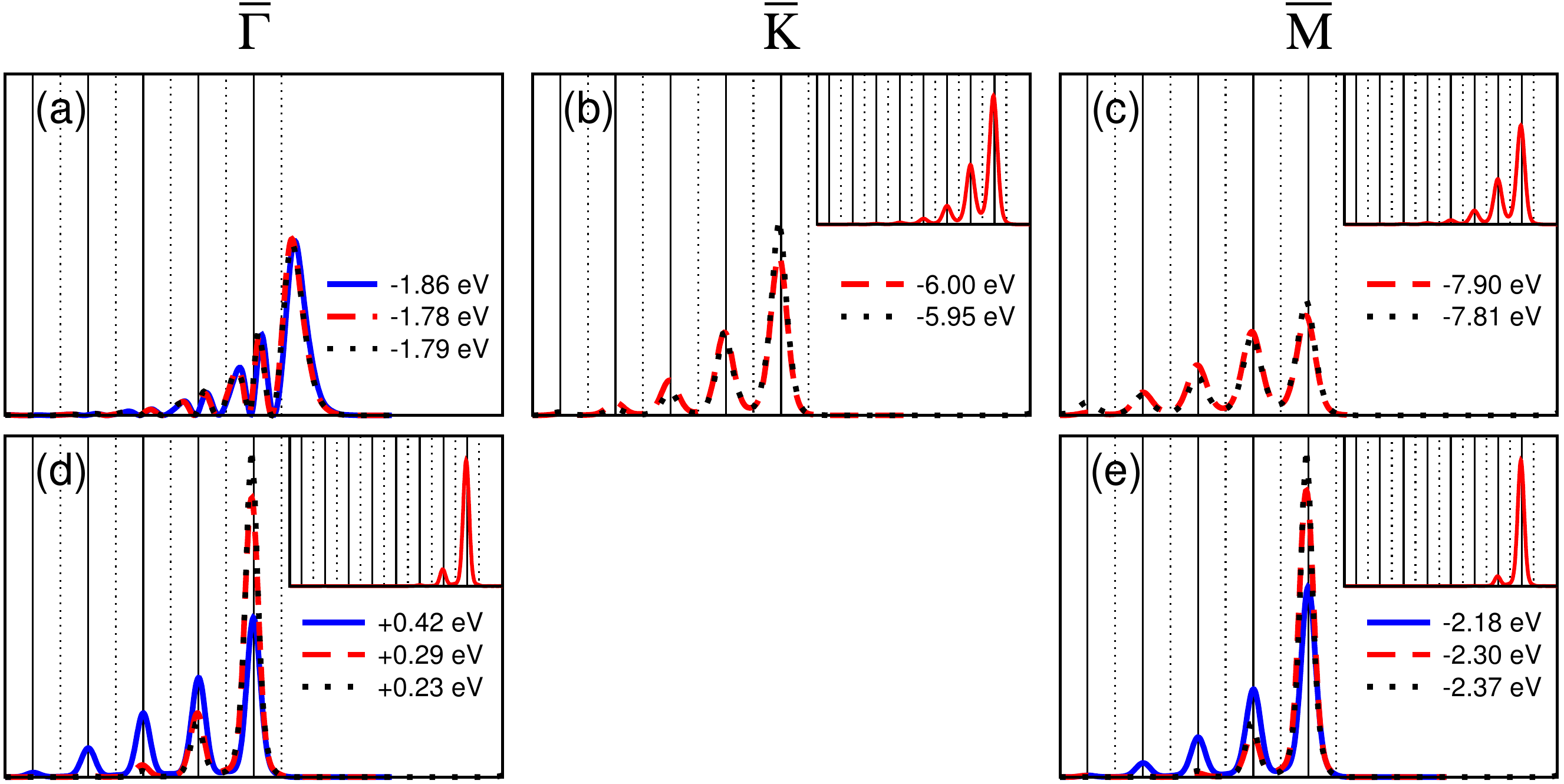}
\caption{(Color online) Projected density of surface and subsurface states \(\varrho_{n,\mathbf{k}}(z)\) in the Mg terminated MgB$_2$(0001) 
surface versus position normal to the surface plane $z$, and Kohn-Sham eigenenergies $\varepsilon_{n,\mathbf{k}}$ relative to the Fermi level $\varepsilon_F$.  Results from plane wave PWscf calculations before ({\color{blue}{\bf{---}}}) and after ({\color{red}{\bf{-- --}}}) structural optimization are compared with real-space GPAW calculations for the relaxed structure ($\cdots$). Solid and dotted vertical lines denote positions of the B and Mg atomic layers, respectively. Inserts show $\varrho_{n,\mathbf{k}}(z)$ from plane wave PWscf calculations for a structurally optimized extended supercell model with six bulk unit cells added in the center of the slab.}
\label{Fig4}
\end{figure*} 

\begin{figure}%[hb]
%\vspace{1cm}
\includegraphics[width=\columnwidth]{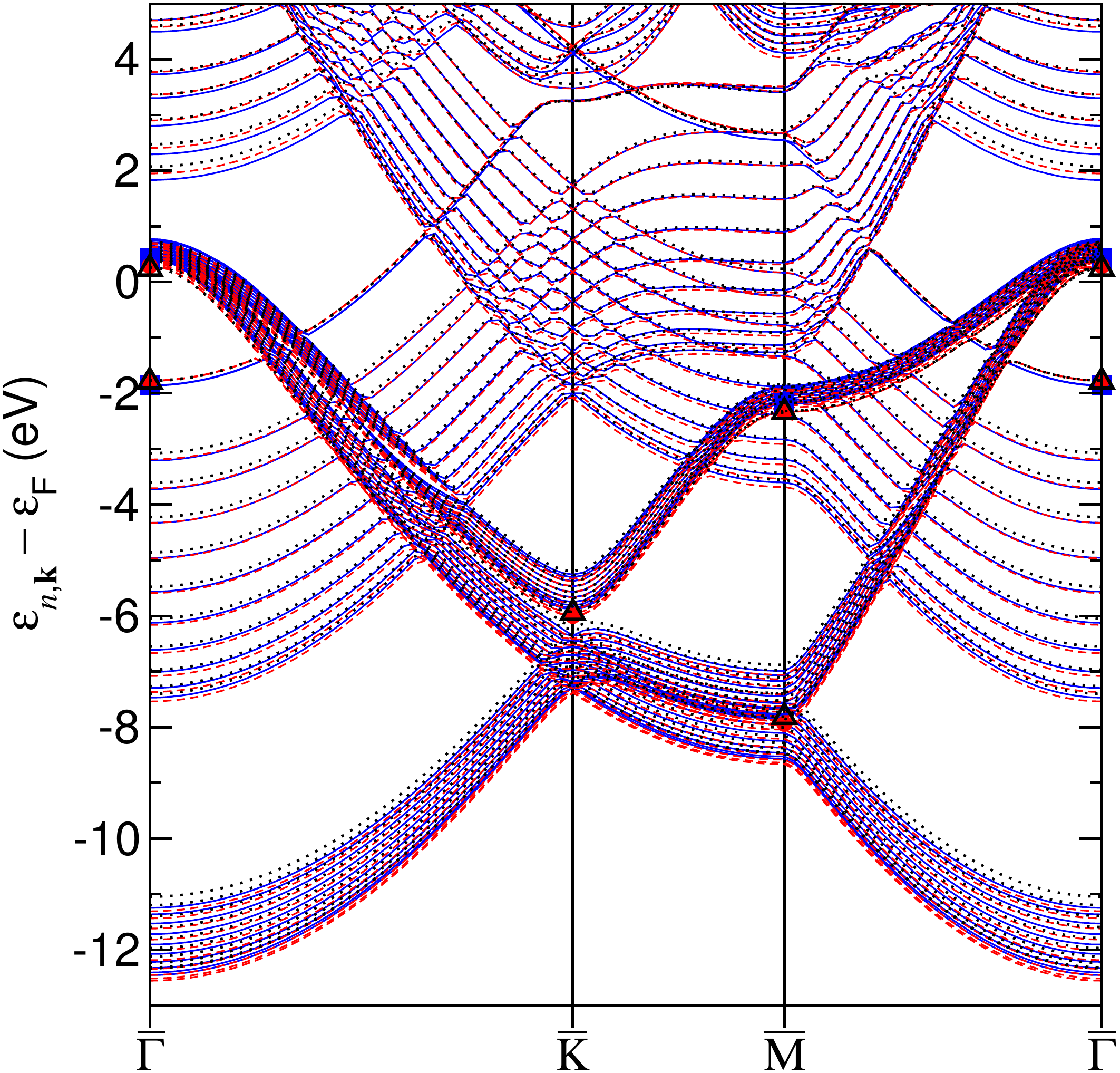}
\caption{(Color online) Band structure (lines) and energies of the surface and subsurface states shown in Fig.~\ref{Fig4} (symbols) for the Mg terminated 
MgB$_2$(0001) surface in eV relative to the Fermi level $\varepsilon_F$. Results from plane wave PWscf calculations before ({\color{blue}{\bf{---}}}; {\color{blue}{$\blacksquare$}}) and after ({\color{red}{\bf{-- --}}}; {\color{red}{$\medbullet$}}) structural optimization are compared with real-space GPAW calculations for the relaxed structure ($\cdots$; $\vartriangle$).}
\label{Fig3}
\end{figure}

\begin{figure}%[h]
%\vspace{2cm}
\includegraphics[width=\columnwidth]{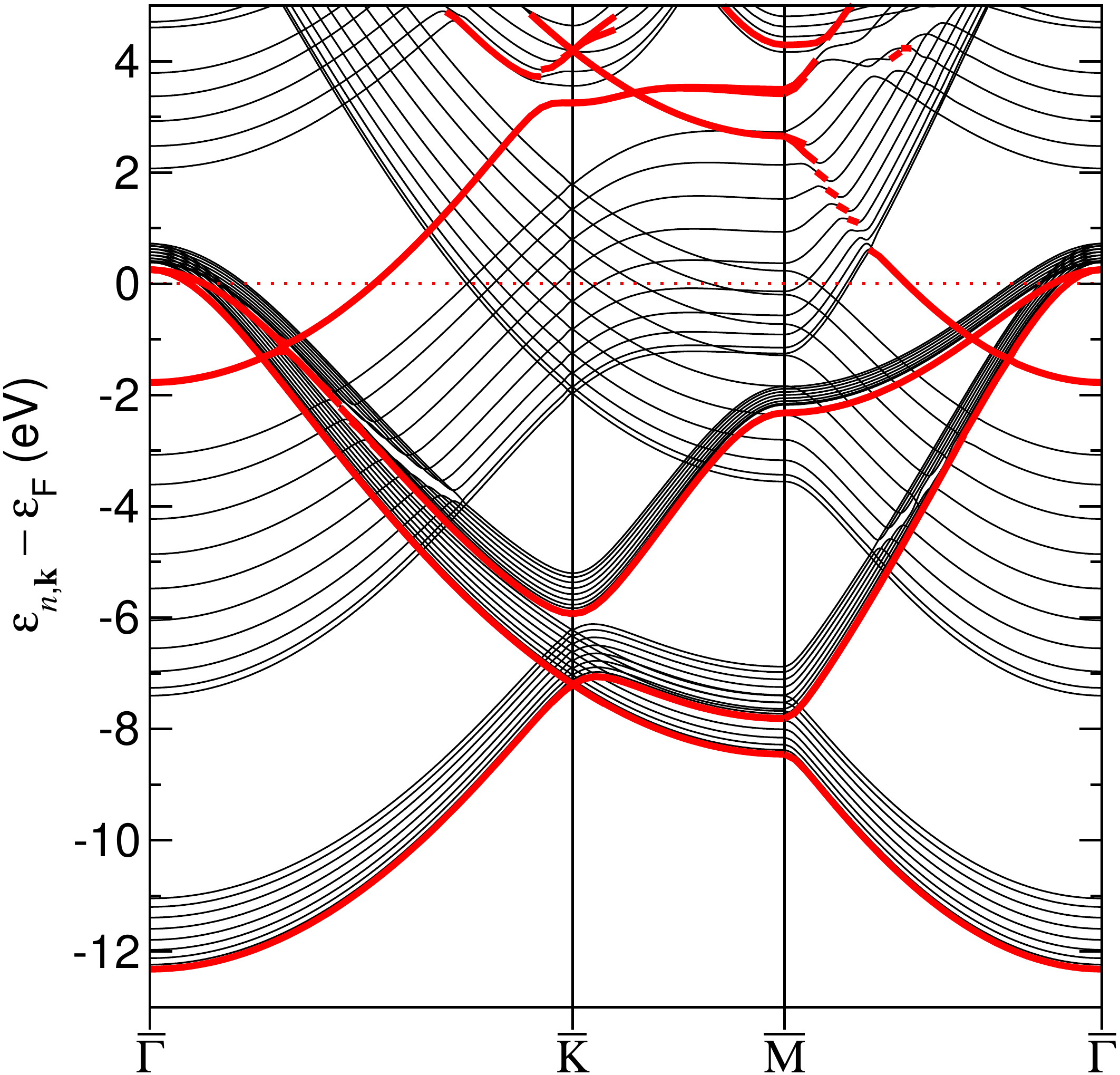}
\caption{(Color online) GPAW overlap-based band structure of relaxed Mg terminated 
MgB$_2$(0001) surface, with \(\psi_{n,\mathbf{k}}\rightarrow\psi_{n',\mathbf{k}+\Delta\mathbf{k}}\) for \(n'\) which maximizes \(\left|\langle\psi_{n,\mathbf{k}}|\psi_{n',\mathbf{k}+\Delta\mathbf{k}}\rangle_{\mathcal{V}}\right|\).  Surface states with more than two thirds of their weight in the first vacuum and first two bulk layers, $s_{n,\mathbf{k}} \gtrsim 0.66$, are shown separately ({\color{red}{\bf{---}}}).}
\label{Fig_GPAW_Mg}
\end{figure}

\begin{table}
\caption{Change in Mg terminated Mg--B inter-layer separation $\Delta_{i,j}$ between the $i^{\mathrm{th}}$ and $j^{\mathrm{th}}$ layers from the surface,  relative to the bulk MgB$_2$ experimental geometry.}\label{Table2}
\begin{ruledtabular}
\begin{tabular}{lrrrr}
&\multicolumn{2}{c}{PWscf LDA\footnote{This Work}}&\multicolumn{2}{c}{VASP PBE\footnote{Ref.~\onlinecite{Li}}}
\\\hline
$\Delta_{0,1}$[Mg--B]& $-5.2\%$&-9.2 pm& $-3.7\%$&-6.5 pm\\
$\Delta_{1,2}$[B--Mg]& $-0.3\%$&-0.5 pm& $ 1.2\%$&  2.1 pm\\
$\Delta_{2,3}$[Mg--B]& $-1.9\%$&-3.3 pm& $ 0.2\%$&  0.4 pm\\
$\Delta_{3,4}$[B--Mg]& $-0.7\%$&-1.2 pm& $ 0.5\%$&  0.9 pm\\
$\Delta_{4,5}$[Mg--B]& $-0.4\%$&-0.7 pm& $-0.3\%$& -0.5 pm\\
$\Delta_{5,6}$[B--Mg]& $-0.9\%$&-1.6 pm&---& ---\\
$\Delta_{6,7}$[Mg--B]& $-0.75\%$&-1.3 pm&---&---\\
$\Delta_{7,8}$[B--Mg]& $-0.9\%$&-1.6 pm&---&---\\
$\Delta_{8,9}$[Mg--B]& $-0.9\%$&-1.6 pm&---&---\\
$\Delta_{9,10}$[B--Mg]& $-0.9\%$&-1.5 pm&---&---\\
\end{tabular}
\end{ruledtabular}
\end{table}

Even though optimization of the crystal structure only slightly modified the Mg terminated surface (\emph{cf.}~Table~\ref{Table2}) and band structure (\emph{cf.}~Fig.~\ref{Fig3}), modifications of the surface electronic structure are quite radical. Density distributions in Fig.~\ref{Fig4} show that for the unrelaxed surface there exist two types of localized states, an $sp_z$ surface 
state (Fig.~\ref{Fig4}(a)) and B $\sigma$ subsurface states (Fig.~\ref{Fig4}(d,e)). On the other hand, even though the relaxed surface is only slightly compressed (maximally $-2.7\%$ in the first layer), it induces the appearance of new surface 
states (Fig.~\ref{Fig4}(b,c)). Such states consist of B $\sigma$ states, which are localized around B layers but 
penetrate quite deep inside the crystal. At the $\overline{\text{K}}$ point this state may even be considered a surface state resonance. Figure \ref{Fig_GPAW_Mg} clearly shows that these states are more properly associated with the B $\sigma_{2}$ and $\sigma_{3}$ surface states (Fig.~\ref{Fig4} (d,e)).

Another consequence of the surface relaxation is that the B $\sigma$ surface states (Fig.~\ref{Fig4}(d,b,e)) are 
pushed upwards towards the surface B layers. In other words, in the relaxed crystal the $\sigma$ surface states are completely localized 
in the one or two topmost B layers. As shown in Fig.~\ref{Fig_GPAW_Mg}, the three B $\sigma$ bands have definite surface character throughout the SBZ, where they follow the bottom edge of the B $\sigma$ bulk bands.  

We also find relaxation lowers the energy of the $\sigma$ subsurface states by approximately $0.1$~eV at the
$\overline{\Gamma}$ and  $\overline{\text{K}}$ points, and increases the energy of the $sp_z$ surface state from $-1.86$~eV to $-1.78$~eV at the $\overline{\Gamma}$ point, in better agreement with the experimental value of $-1.6$~eV.  It should also be noted that the B $\sigma_{3}$ surface band is more localized on the topmost B layer in the real-space GPAW calculation.  This is attributable to the inherit difficulties plane wave based methods have in describing state localization, compared to real-space methods.

In fact, from Fig.~\ref{Fig_GPAW_Mg} we see that the bottom of all three B $\sigma$ bands have $s_{n\mathbf{k}} \gtrsim$ 0.66 throughout the SBZ.  This suggests that adding a layer of Mg shifts the B $\sigma$ surface bands down in energy by about 0.9 eV relative to a the B terminated MgB$_2$(0001) surface.  On the other hand, we find that the $sp_z$ surface band is higher up in energy by about 0.9 eV relative to the B terminated surface.  In Sec.~\ref{Discuss} we will discuss how this rigid shifting in energy of the surface bands may be understood in terms of charging of the MgB$_2$(0001) surface.  

\subsection{Li-terminated MgB$_{\bf 2}$(0001) surface}

\begin{figure*}
\includegraphics[width=0.7\textwidth]{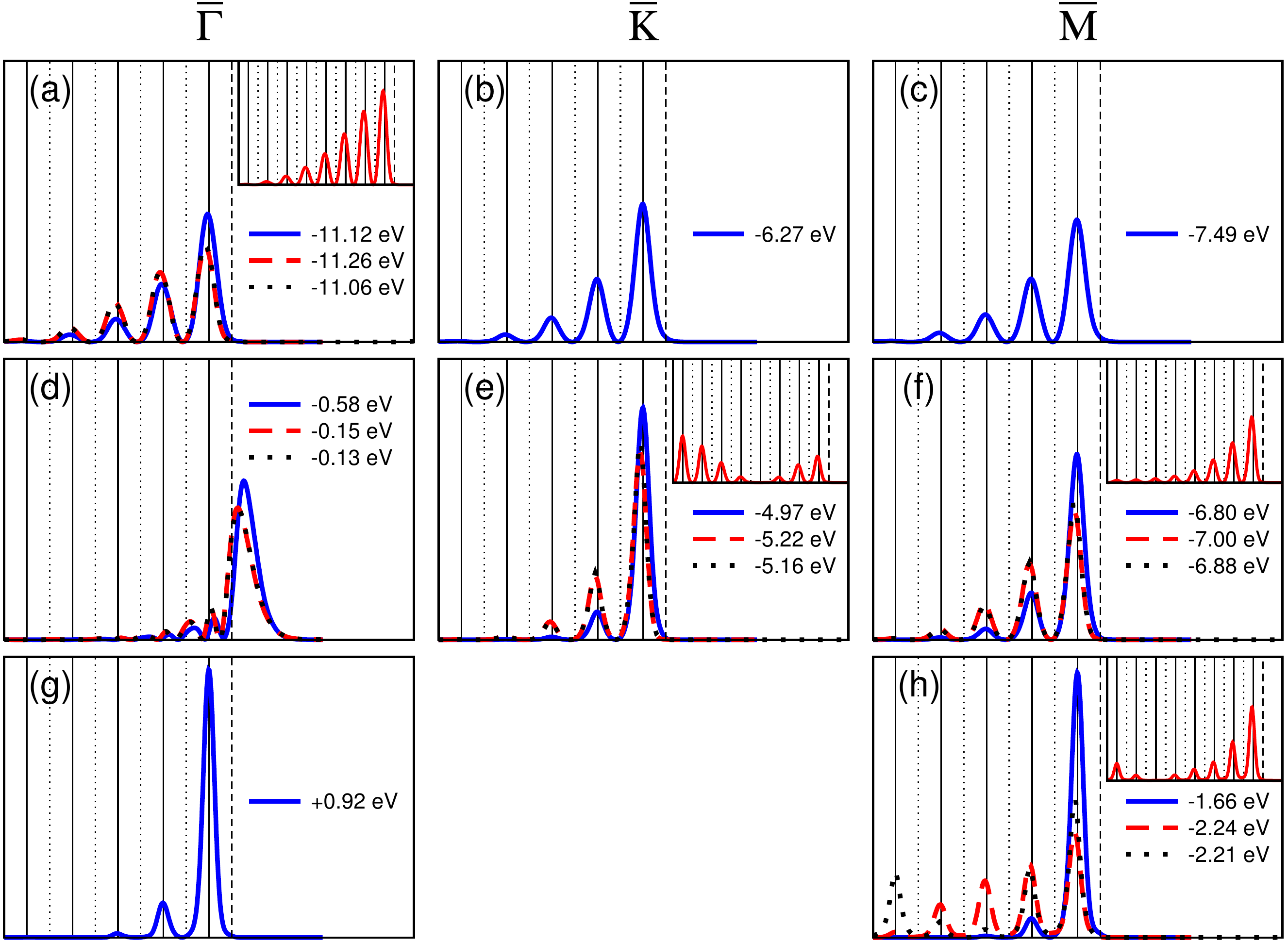}
\caption{(Color online) Projected density of surface and subsurface states \(\varrho_{n,\mathbf{k}}(z)\) in the Li terminated MgB$_2$(0001) 
surface versus position normal to the surface plane $z$, and Kohn-Sham eigenenergies $\varepsilon_{n,\mathbf{k}}$ relative to the Fermi level $\varepsilon_F$.  Results from plane wave PWscf calculations before ({\color{blue}{\bf{---}}}) and after ({\color{red}{\bf{-- --}}}) structural optimization are compared with real-space GPAW calculations for the relaxed structure ($\cdots$). Solid, dotted, and dashed vertical lines denote positions of the B, Mg, and Li atomic layers, respectively. Inserts show $\varrho_{n,\mathbf{k}}(z)$ from plane wave PWscf calculations for a structurally optimized extended supercell model with six bulk unit cells added in the center of the slab.}
\label{Fig6}
\end{figure*} 

\begin{figure}
%\vspace{2cm}
\includegraphics[width=\columnwidth]{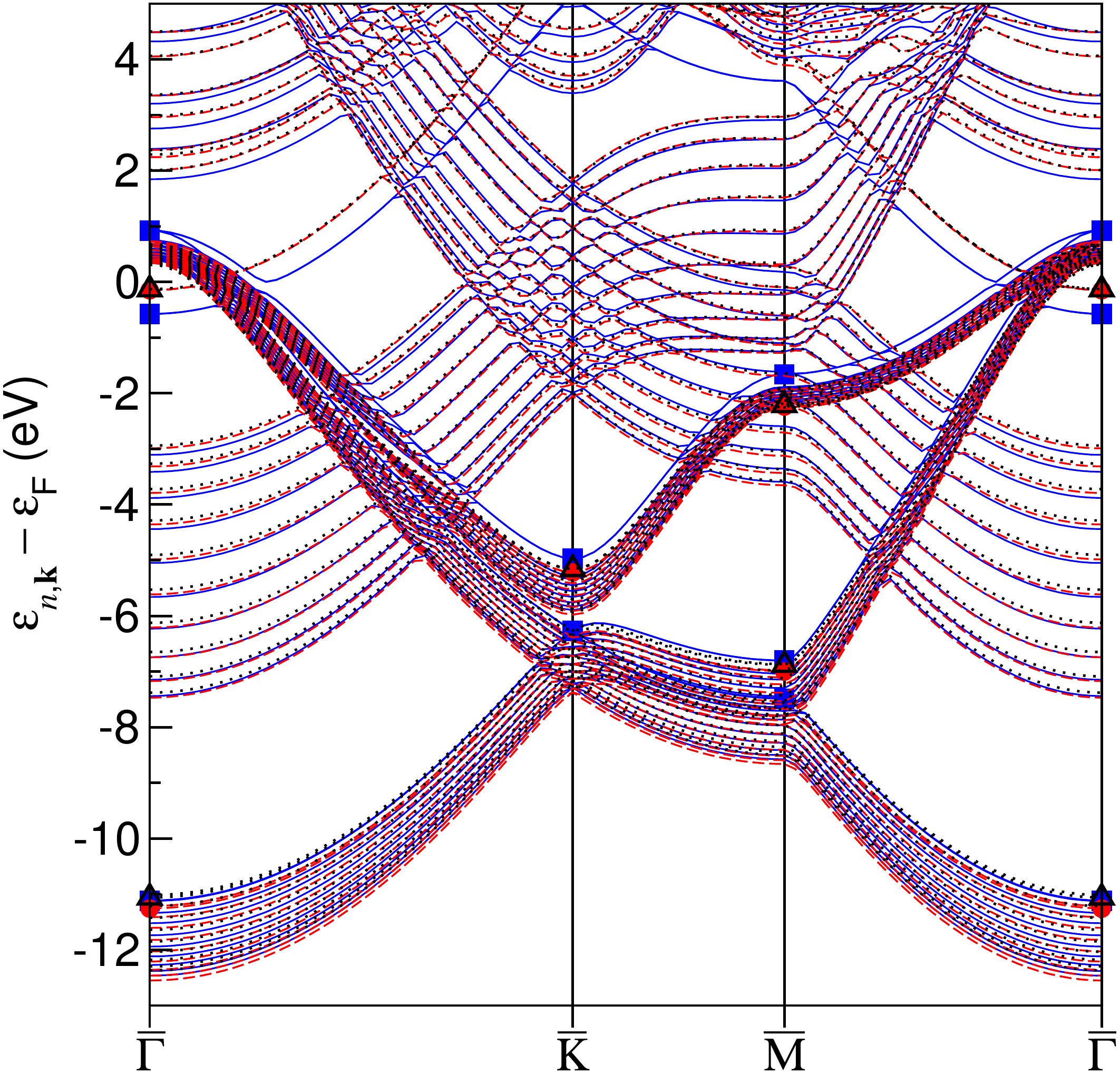}
\caption{(Color online) Band structure (lines) and energies of the surface and subsurface states shown in Fig.~\ref{Fig6} (symbols) for the Li terminated 
MgB$_2$(0001) surface in eV relative to the Fermi level $\varepsilon_F$. Results from plane wave PWscf calculations before ({\color{blue}{\bf{---}}}; {\color{blue}{$\blacksquare$}}) and after ({\color{red}{\bf{-- --}}}; {\color{red}{$\medbullet$}}) structural optimization are compared with real-space GPAW calculations for the relaxed structure ($\cdots$; $\vartriangle$).}
\label{Fig5}
\end{figure} 

\begin{figure}
%\vspace{2cm}
\includegraphics[width=\columnwidth]{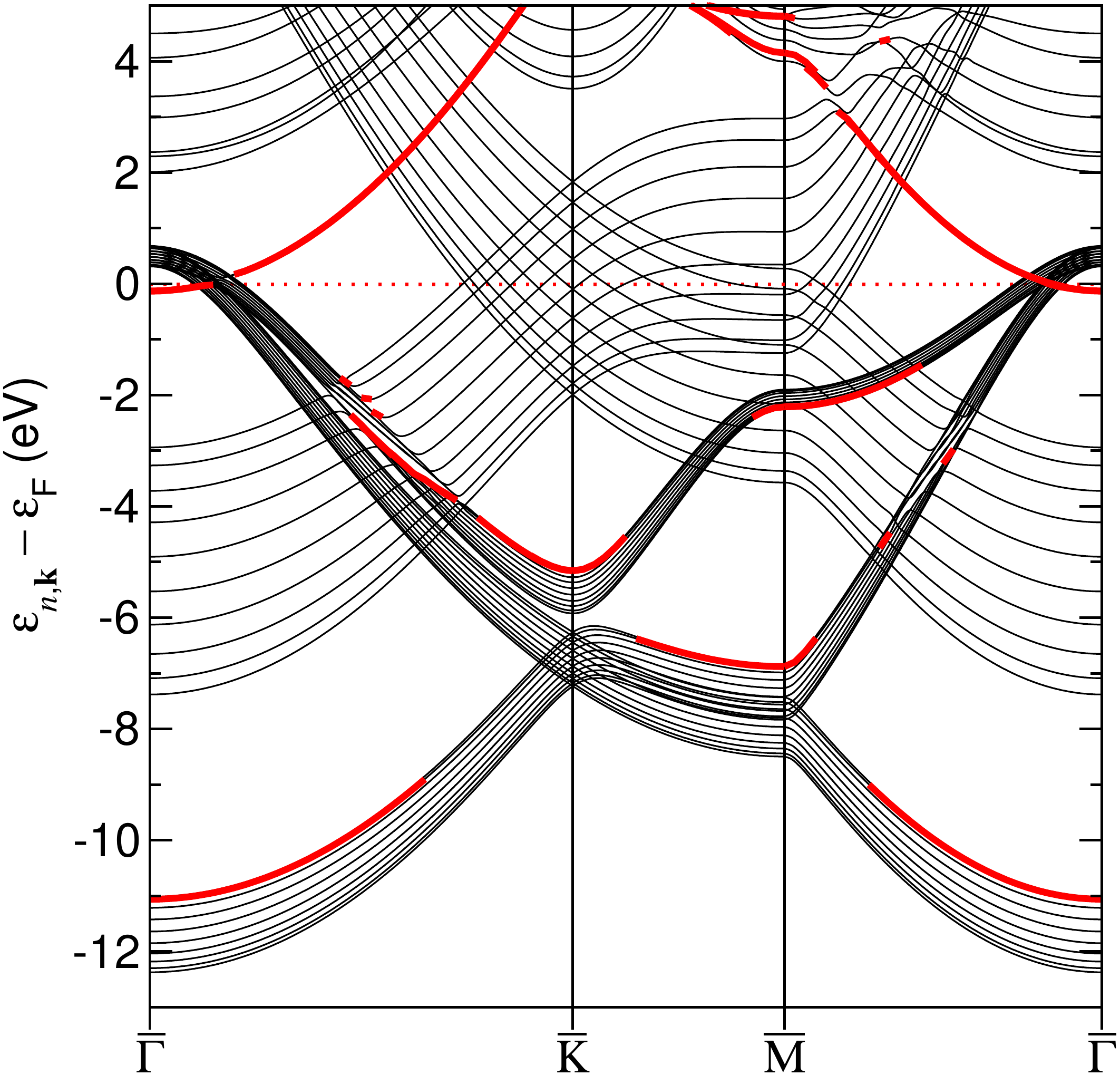}
\caption{(Color online) GPAW overlap-based band structure of relaxed Li terminated 
MgB$_2$(0001) surface, with \(\psi_{n,\mathbf{k}}\rightarrow\psi_{n',\mathbf{k}+\Delta\mathbf{k}}\) for \(n'\) which maximizes \(\left|\langle\psi_{n,\mathbf{k}}|\psi_{n',\mathbf{k}+\Delta\mathbf{k}}\rangle_{\mathcal{V}}\right|\).  Surface states with more than two thirds of their weight in the first vacuum and first two bulk layers, $s_{n,\mathbf{k}} \gtrsim 0.66$, are shown separately ({\color{red}{\bf{---}}}).}
\label{Fig_GPAW_Li}
\end{figure} 

\begin{table}
\caption{Change in Li terminated Mg--B inter-layer separation $\Delta_{i,j}$ between the $i^{\mathrm{th}}$ and $j^{\mathrm{th}}$ layers from the surface,  relative to the bulk MgB$_2$ experimental geometry.}\label{Table3}
\begin{ruledtabular}
\begin{tabular}{lrr}
&\multicolumn{2}{c}{PWscf LDA\footnote{This Work}}
\\\hline
$\Delta_{0,1}$[Li--B]& $-29.8\%$&-52.4 pm\\
$\Delta_{1,2}$[B--Mg]& $-0.6\%$&-1.0 pm\\
$\Delta_{2,3}$[Mg--B]& $-1.2\%$&-2.2 pm\\
$\Delta_{3,4}$[B--Mg]& $-1.5\%$&-2.7 pm\\
$\Delta_{4,5}$[Mg--B]& $-1.2\%$&-2.1 pm\\
$\Delta_{5,6}$[B--Mg]& $-1.5\%$&-2.6 pm\\
$\Delta_{6,7}$[Mg--B]& $-1.4\%$&-2.4 pm\\
$\Delta_{7,8}$[B--Mg]& $-1.9\%$&-3.3 pm\\
$\Delta_{8,9}$[Mg--B]& $-1.5\%$&-2.6 pm\\
$\Delta_{9,10}$[B--Mg]& $-1.5\%$&-2.7 pm\\
\end{tabular}
\end{ruledtabular}
\end{table}

To understand how surface charging, or electro-chemical doping might affect MgB$_2$(0001) surface states, we have directly replaced Mg by Li in the first atomic layer (Mg$\rightarrow$Li).  In effect, this approximates the removal of one electron from each of the Mg terminated surfaces. In other words, replacing Mg by Li resembles a positively charged Mg terminated (Mg$^{+1}$) surface. This allows us to directly probe the effect of surface charging of the Mg terminated surface without the need for compensating background charges in the calculation.  We have then performed structural optimization of the Li terminated MgB$_2$(0001) surface to model how Li termination changes the surface states.  From Table~\ref{Table3} we see that the smaller atomic radius of Li induces a substantial relaxation, with the relaxed Li atomic layer moved down by 52.4 pm towards the B atomic layer.  

Both the surface charging and the crystal structure modification strongly change the electronic structure of the surface states. Figure \ref{Fig5} shows that the lowest B $\sigma_1$ surface resonance band for the structurally unoptimized surface (Mg$\rightarrow$Li$\sim$Mg$^{+1}$), is slightly separated from the 
upper edge of the bulk B $\sigma_1$ band. The density distributions of such states are shown in Fig.~\ref{Fig6}(a,b,c,f), where we see that B $\sigma_1$ surface resonances are mainly localized in the four topmost B layers. 
Structural optimization causes a slight downward shift of the surface resonance B $\sigma_1$ bands and induces a strong interaction/hybridization with the bulk B $\sigma_1$ band, so that they decay into the bulk states. I.e., it increases the amplitude of the propagation of the resonance state into the bulk, as can be seen at the
$\overline{\Gamma}$ point in Fig.~\ref{Fig6}(a). For $\overline{\text{K}}$ and $\overline{\text{M}}$ points, the B $\sigma_1$ surface resonances cannot even be distinguished from the normal bulk states and are not 
shown.  This is seen clearly in Fig.~\ref{Fig_GPAW_Li}, with the relaxed Li terminated B $\sigma_1$ surface states becoming increasingly bulk-like between $\overline{\Gamma}$ and K, until $s_{n,\mathbf{k}} < 0.66$.  

Relaxation also causes a $0.4$~eV reduction of the binding energy of the $sp_z$ supra-surface state band. This is expected 
because relaxation causes contraction of the topmost Li-B interstitial region (\emph{cf.}~Table \ref{Table3}) where the $sp_z$ surface state is mostly located (\emph{cf.}~Fig.~\ref{Fig6}).  
The binding energy of the $sp_z$ surface state at $\overline{\Gamma}$ point is strongly reduced, from about $2.6$~eV to $0.1$~eV, comparing with the B terminated 
surface. This suggests that the binding energy of the $sp_z$ surface state at the $\overline{\Gamma}$ point may be strongly influenced by surface contamination or doping of the B terminated surface, as we will discuss in Sec.~\ref{Discuss}.

We also find the more localized B $\sigma_2$ and $\sigma_3$ bands are strongly modified by Li termination. As can be seen in Fig.~\ref{Fig5} for Mg$\rightarrow$Li$\sim$Mg$^{+1}$, the B $\sigma_{2}$ and $\sigma_3$ surface bands are always 
slightly separated from the upper edge of the bulk B $\sigma_{2}$ and $\sigma_3$ bands, with their density highly localized in the first one or two B layers, as seen in Fig.~\ref{Fig6}(g,e,h). These states resemble the highly localized B $\sigma_{2}$ and $\sigma_3$ surface states in the B terminated surface.     
However, surface relaxation shifts the localized B $\sigma_2$ and $\sigma_3$ bands down in energy.  This causes a strong interaction between localized and bulk $\sigma$ states, with the surface state decaying into the bulk states. This is clearly illustrated by the densities of these states for the relaxed Li terminated structure shown in Fig.~\ref{Fig6}(e,h). Here we see that the $\sigma$ localized state now penetrates substantially into the crystal. By plotting the 
same distributions for an extended crystal structure (\emph{cf.}~insets of Fig.~\ref{Fig6}(e,h)) we see that 
these states are actually localized state resonances or even pure bulk states at the $\overline{\text{K}}$ point.  From Fig.~\ref{Fig_GPAW_Li} we see that the B $\sigma_{2}$ and $\sigma_3$ surface states strongly hybridize with the bulk $\sigma$ states, with $s_{n,\mathbf{k}} < 0.66$ over much of the SBZ.

From this we make two observations. First, removing charge from the Mg terminated surface would strongly affect the surface and subsurface states. Second, for the relaxed Li terminated surface all subsurface states hybridize into localized state resonances or bulk states, while the $sp_z$ surface state survives. We will discuss how these affects may be understood and controlled, with the goal of tuning the surface states of MgB$_2$, in the following section. 

\section{Discussion}
\label{Discuss}

\begin{figure}
\includegraphics[width=0.85\columnwidth]{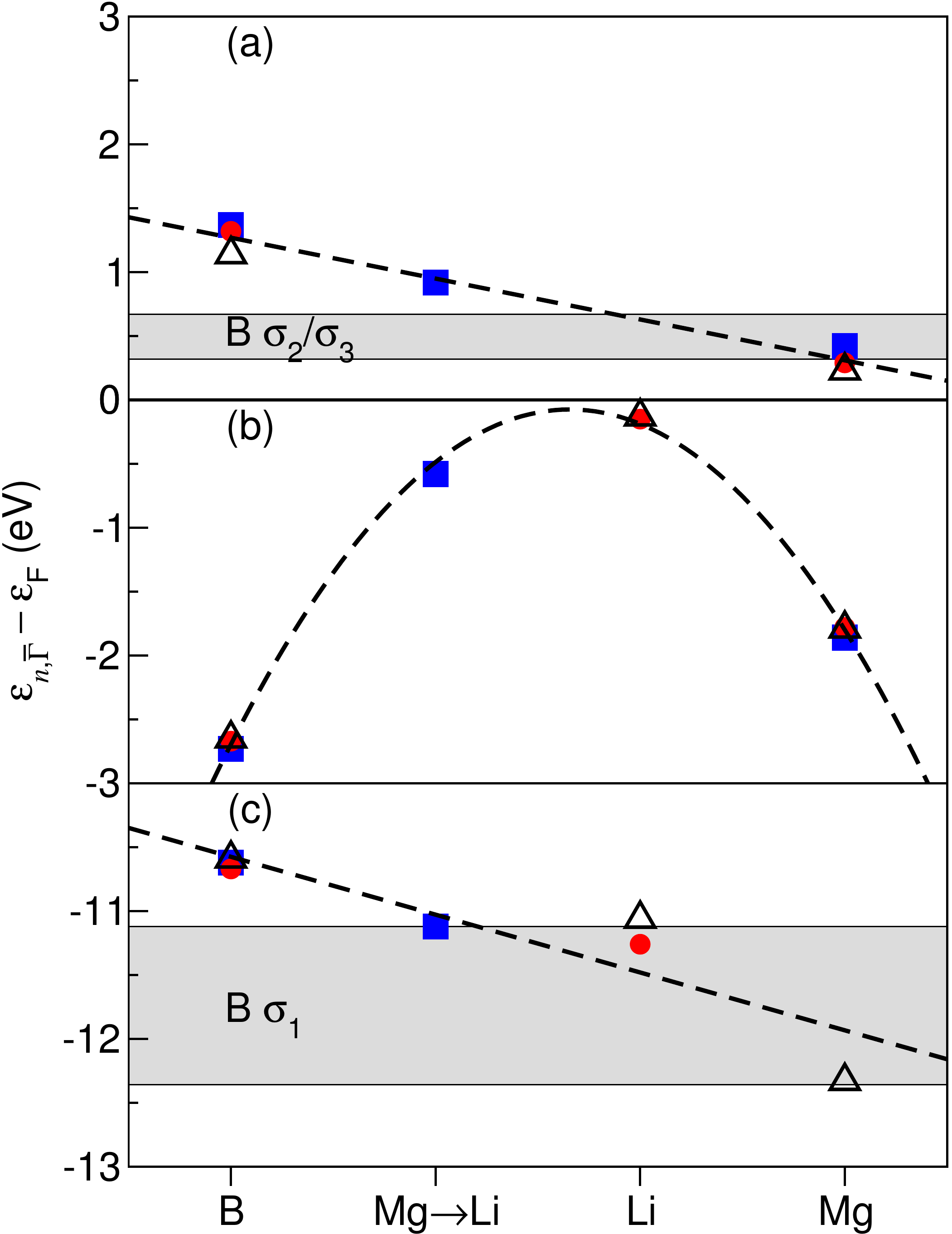}
\caption{Surface and subsurface state energies at the $\overline{\Gamma}$ point, \(\varepsilon_{n,\overline{\Gamma}}\) versus surface termination and doping of the MgB$_2$(0001) surface by B, Mg$\rightarrow$Li$\sim$Mg$^{+1}$, Li, and Mg, in eV relative to the Fermi level $\varepsilon_F$, for (a) B $\sigma_{2}$ and $\sigma_3$, (b) $sp_z$, and (c) B $\sigma_1$ states.  Results from plane wave PWscf calculations before ({\color{blue}{$\blacksquare$}}) and after ({\color{red}{$\medbullet$}}) structural optimization are compared with real-space GPAW calculations for the relaxed structures ($\vartriangle$). The B $\sigma_{2,3}$ and B $\sigma_1$ bulk band regions are shaded gray.  Fits to the surface state positions ({\bf{-- --}}) are provided as guides to the eye.}\label{Trends}
\end{figure}

Comparing the relative positions of the B $\sigma$ and $sp_z$ surface states for B, Mg, and Li terminations, shown in Figs.~\ref{Fig_GPAW_B}, \ref{Fig_GPAW_Mg}, and \ref{Fig_GPAW_Li}, respectively, clear trends emerge.  These suggest that changing surface termination or doping electro-chemically can tune both the energy and localization of the MgB$_2$(0001) surface states.

To demonstrate this point, we replot in Fig.~\ref{Trends} the energies of the surface and subsurface states at the $\overline{\Gamma}$ point as a function of surface termination. As we move from left to right in Fig.~\ref{Trends}, charge is transferred to the B atomic layer, as we move from neutral (B), to charging of less than $-1 e$ (Mg$\rightarrow$Li$\sim$Mg$^{+1}$), of about $-1 e$ (Li), and finally, of about $-2 e$ (Mg).  As charge is donated to the B atomic layer, both the degenerate B $\sigma_{2}$ and $\sigma_3$ levels (shown in Fig.~\ref{Trends}(a)) and B $\sigma_1$ levels (shown in Fig.~\ref{Trends}(c)) are filled and move down in energy relative to the bulk MgB$_2$ Fermi level, in a quasi-linear fashion.  Further, when surface states cross the bulk B $\sigma$ bands, they hybridize, and lose much of their weight on the surface.  This is clearly seen in Fig.~\ref{Fig_GPAW_Li}, where for the B $\sigma_{2}$ and $\sigma_3$ states $s_{n,\overline{\Gamma}} < 0.66$, and the B $\sigma_1$ state shown in Fig.~\ref{Fig6}(a), also has more weight in the bulk.  This strongly suggests that by electro-chemically doping the surface, we may tune both the energy and localization of the B $\sigma$ surface states.

For the higher energy B $\sigma_{2}$ and $\sigma_3$ surface states this is particularly interesting, as they resemble quantum well 
(QW) states.  This is because the occupying electrons effectively feel the potential of a QW centered around the B 
layer. Moreover, electrons in such states are trapped in the topmost B layer but are ``free'' to move within the band. In this way, the B 
$\sigma$  band may be considered a 2D electron gas decoupled from the surrounding 3D electron gas of the bulk MgB$_2$.       
Such a two-component electron plasma allows different types of plasma dispersions relations. For example, a noble acoustic plasmon may be present, whose 
dispersion resembles that of the acoustic plasmons. Such a plasmon may affect (screen) the electron-phonon coupling.  

On the other hand, for the $sp_z$ state shown in Fig.~\ref{Trends}(b), the energy of the surface state does not correlate with charging of the B atomic layer.  Instead, to understand the behavior of the $sp_z$ state, we must first consider its projected density $\varrho_{n,\overline{\Gamma}}(z)$.   For each termination we find that the $sp_z$ state is an ionic/metallic bonding state between the Mg/Li and B atomic layers. This is clear from Figs.~\ref{Fig2}(d), \ref{Fig4}(a), and \ref{Fig6}(d), where we see that the $sp_z$ state has mostly $p_z$ character localized in the inter-layer region.  Specifically, for B termination the $sp_z$ has significant weight in the bulk, while for Mg and Li terminations the $sp_z$ state has more $s$ character, with significant weight in the vacuum region. 

This is also clear from the band structures shown in Figs.~\ref{Fig_GPAW_B}, \ref{Fig_GPAW_Mg}, and \ref{Fig_GPAW_Li}.  Specifically, for B termination (\emph{cf.}~Fig.~\ref{Fig_GPAW_B}) the $sp_z$ state is at the top of the broad $\pi$ band, having mostly $p_z$ character, and is filled throughout most of the SBZ.  For the Mg terminated surface (\emph{cf.}~Fig.~\ref{Fig_GPAW_Mg}) the $sp_z$ state has more $s$ character, is separated from the bulk $\pi$ bands although still following the $\pi$ bands through the SBZ, is shifted up in energy, and partially emptied.  When we directly replace Mg by Li to model a Mg$^{+1}$ surface, we find that the $sp_z$ surface state has mostly Mg/Li $s$ character (\emph{cf.}~Fig.~\ref{Fig6}(d)), is shifted up in energy between the $\pi$ and Mg $s$ bulk states, and is almost completely empty.  For the relaxed Li terminated surface (\emph{cf.}~Fig.~\ref{Fig_GPAW_Li}) the $sp_z$ state is nearly emptied, and has significant $s$ character, as shown in Fig.~\ref{Fig6}(d).

Together, this suggests that by altering the termination of the MgB$_2$(0001) surface, the $sp_z$ surface state is changed from a bulk-like state with $p_z$ character for B termination, to an increasingly localized and emptied surface state with more metallic $s$ character from Mg termination, to Mg$\rightarrow$Li (or Mg$^{+1}$) termination, and finally the structurally optimized Li termination.   This is consistent with previous calculations for the Na terminated MgB$_2$(0001) surface \cite{Profeta}, which found the $sp_z$ surface state between that calculated here for Mg and Mg$\rightarrow$Li terminations.  In effect, by varying the surface termination or electro-chemically doping the surface, one can tune both the localization and energy range of MgB$_2$(0001) $sp_z$ surface states.  

In this way, one could then modify the density or Fermi velocity of the charge carriers on the 
surface and consequently surface plasmons dispersions. This means that surface plasmon characteristics may be tuned rather easily by surface charging or doping.   

\section{Conclusions}
\label{Conclus}
In this paper we have investigated ​MgB$_2$(0001) ​surface and subsurface states, and the influence of surface termination (by B, Mg, and Li) and surface charging (replacing Mg by Li to model Mg$^{+1}$) using both plane wave and real-space {\em ab initio} methods.
Generally we find for the MgB$_2$(0001) surface there exist three types of surface states which are often well separated from bulk bands, namely, the B $\sigma_1$ surface band, the $sp_z$ supra-surface state band, and the B $\sigma_{2}$ and $\sigma_e$ QW bands. 

If charge is added to the B layer by either changing the surface termination or surface charging, the B $\sigma$ surface bands shift shift down in energy.  As these bands cross the bulk B $\sigma$ bands, they lose their surface state character due to hybridization.  In effect, by adjusting the charging of the B subsurface atomic layer, we may tune both the localization and energy range of the B $\sigma$ surface states.  

On the other hand, the addition of a Mg, Mg$^{+1}$, or Li layer above the B surface layer gives the $sp_z$ surface state band an increasingly metallic $s$ character as it is emptied.  We find these results are quite robust, and independent of the \emph{ab initio} methodology employed (plane wave or real-space) and relaxation of the surface. 

Overall, these results suggest that by changing surface coverage and charging, one may tune both the density and Fermi velocity of the charge carriers at the surface. This would make possible the controlled tuning of the surface plasmon
dispersions, e.g. the slope of the acoustic surface plasmon, which may
then influence (modify or enhance) ``surface'' superconductivity.  This has profound implications for both the superconducting behaviour and surface or acoustic surface plasmon dispersion of MgB$_2$. 

Therefore, because of the diversity of MgB$_2$'s surface electronic structure, 
with combinations of various types of quasi-2D and 3D plasmas, it seems that the MgB$_2$ surface should have very unusual dielectric and 
optical properties. To clarify these unique properties, our next step will be the investigation of the MgB$_2$ surface dielectric response. 

\acknowledgments
We acknowledge funding through the Spanish ``Juan de la Cierva'' program (JCI-2010-08156), Spanish Ministerio de Ciencia e Innovac{\'{\i}}on (FIS2010-21282-C02-01, FIS2010-19609-C02-01), Spanish ``Grupos Consolidados UPV/EHU del Gobierno Vasco'' (IT-319-07, IT-366-07), and ACI-Promociona (ACI2009-1036). The European Theoretical Spectroscopy Facility is funded through ETSF-I3 (Contract Number 211956).
%\bibliography{../../bibliography}
%\bibliographystyle{arXiv}

\end{document}